# MEASURING THE CKM MATRIX ELEMENT $V_{tb}$ AT DØ AND CDF

A.P. HEINSON

*Department of Physics, University of California, Riverside, CA 92521*

I present measurements by the CDF collaboration of the Standard Model three generation CKM matrix element $V_{tb}$ and of a special case extension with additional assumptions, using current Tevatron $t\bar{t}$ data. I then show how we can significantly improve the precision on $V_{tb}$ and at the same time extend the measurement so it is not constrained by Standard Model assumptions, using single top production at the upgraded Tevatron.

## 1 Introduction

By convention, the charge $+2/3$ quarks ($u$, $c$, and $t$) are unmixed, and mixing of the charge $-1/3$ quarks is described by the $3\times 3$ unitary Cabibbo-Kobayashi-Maskawa (CKM) matrix[1] $V$:

$$\begin{pmatrix} d' \\ s' \\ b' \end{pmatrix} = \begin{pmatrix} V_{ud} & V_{us} & V_{ub} \\ V_{cd} & V_{cs} & V_{cb} \\ V_{td} & V_{ts} & V_{tb} \end{pmatrix} \begin{pmatrix} d \\ s \\ b \end{pmatrix}.$$

If we assume unitarity and only three generations of quarks, then from direct and indirect measurements of the other elements, the magnitude of $V_{tb}$ is very precisely known: $|V_{tb}| = 0.9991 \pm 0.0001$, without any direct measurements using the top quark system. However, once these two constraints are removed, then the allowed 90% confidence level (CL) limits[2] open up to leave essentially no bounds on $V_{tb}$:

$$\begin{pmatrix} - & - & - \\ - & - & - \\ 0.004 \rightarrow & 0.034 \rightarrow & 0.9989 \rightarrow \\ 0.014 & 0.046 & 0.9993 \end{pmatrix}_{\text{3 gen}} \Rightarrow \begin{pmatrix} - & - & - & \cdots \\ - & - & - & \cdots \\ 0 \rightarrow & 0 \rightarrow & 0 \rightarrow & \cdots \\ 0.11 & 0.52 & 0.9993 & \\ \vdots & \vdots & \vdots & \ddots \end{pmatrix}_{\text{4+ gen}}.$$

This paper describes a direct measurement of $V_{tb}$ in the three generation Standard Model (SM) using top quark decays in $t\bar{t}$ events, and then presents a much more powerful technique for making a less constrained measurement by DØ and CDF at the upgraded Tevatron using single top production.

---

## 2 Measuring $V_{tb}$ with Current Data

### 2.1 Using $t\bar{t}$ Decays to Measure $V_{tb}$

At the Tevatron collider, a $p\bar{p}$ machine operating at $\sqrt{s} = 1.8$ TeV, top quarks are produced mainly in pairs, with ~90% coming from $q\bar{q} \to t\bar{t}$ and the remaining ~10% from gluon fusion. The DØ and CDF experiments collected integrated luminosities of 125 pb$^{-1}$ and 110 pb$^{-1}$ respectively during the last run, 1992–1996. These two data sets contain between them about 74 reconstructed $t\bar{t}$ events (after background subtraction) with one or more identified leptonically decaying $W$ bosons. Decays of the top quarks and $\bar{t}$ antiquarks can be used to measure $V_{tb}$ because the $tbW$ vertex contains this CKM matrix element:

$$\Gamma = \frac{eV_{tb}}{2\sqrt{2}\sin\theta_w}\left[\gamma_\mu\left(1-\gamma_5\right)\right]$$

(shown in the unitarity gauge). Here, $\theta_w$ is the weak mixing angle, and $\gamma_\mu$ and $\gamma_5$ are Dirac matrices.

To measure $V_{tb}$, one must first assume that the top quark decays to a $W$ boson and a quark (and not a charged Higgs boson or a squark, for instance), then that there are three and only three quark generations, and also that unitarity holds. These assumptions produce a measurement of $V_{tb}$ in the SM. One measures the ratio of branching fractions of top decaying to a $b$ quark to top decaying to any down-type quark $Q$:

$$R \equiv \frac{B(t \to Wb)}{B(t \to WQ)} = \frac{|V_{tb}|^2}{|V_{td}|^2 + |V_{ts}|^2 + |V_{tb}|^2} = \frac{|V_{tb}|^2}{1}.$$

If, however, there are more than three generations of quarks, then the unitarity constraint becomes $|V_{td}|^2 + |V_{ts}|^2 + |V_{tb}|^2 \leq 1$ and we can no longer measure $V_{tb}$ without additional assumptions.

The CDF collaboration has recently presented a new result using their entire data set,[3] and I discuss this measurement here. DØ has a similar analysis in progress which will be completed in the near future.

### 2.2 The CDF Measurement of $V_{tb}$

The CDF collaboration has used their full data set to measure $R$ and hence $V_{tb}$, and has obtained results for the SM and for a special non-SM case. They use both single-lepton + ≥4 jets events ("$l$+jets") and dilepton events, where the leptons are from $W$ decays from $t \to WQ$. The method is as follows. If one could obtain a pure $t\bar{t}$ sample, then $R$ would be:

$$R \equiv \frac{B(t \to Wb)}{B(t \to WQ)} = \frac{\text{No. of tagged events}/\varepsilon_b}{\text{No. of untagged events} + \text{No. of tagged events}/\varepsilon_b},$$

where $\varepsilon_b$ is the efficiency for tagging a $b$ jet. However, since there are backgrounds in all the event samples, a more sophisticated method is needed. The data is divided into four orthogonal bins, defined below, and a likelihood is constructed which is a function of the numbers of observed and background events in each of these four bins, and of $\varepsilon_b$. The value of $R$ is then found which maximizes the likelihood, and hence one obtains a measurement of $V_{tb}$. For this analysis, a $b$ jet is identified using two methods: (i) the SVX silicon vertex detector is used to find a detached secondary vertex in the jet from the long lifetime of the $B$ hadron; and (ii) SLT, or "soft lepton tagging", where an electron or muon is sought close to or in the jet from a semileptonic decay of the $B$ hadron or its charmed daughter. There are 163 $l$+jets and 9 dilepton events, divided into four orthogonal bins as follows. The $l$+jets sample has 126 events with no $b$ tags – mainly background, 14 events with one or more SLT tags and no SVX tags, 18 events with exactly one SVX tagged jet, and 5 events with two SVX tags. The dilepton sample has 6 events with no tags, 3 events with exactly one SVX tag, and no events in the other two bins. SLT tags are ignored when there is an SVX tag, because the mistag rate is ~2% per jet for the SLT technique, but only ~0.5% per jet for the SVX method.[4]

It is important to understand the $b$ tagging efficiency well for this measurement. Other experimental quantities such as trigger efficiencies, acceptances, and particle identification cuts cancel in the ratio $R$, but $\varepsilon_b$ does not as it is convolved with each quantity separately. From the equation above, one can see that the sensitivity to measure $R$ is dependent not on the absolute value of $\varepsilon_b$, but on how small one can make its error. The tagging efficiencies achieved by CDF are $(30.5 \pm 3.0)\%$ per jet using the SVX method, and $(10 \pm 1)\%$ per jet for the SLT technique. The final precision on $V_{tb}$ is currently limited by the event statistics for measuring $\Delta \varepsilon_b$.

The following results are preliminary. For the SM:

$$R = 0.99 \pm 0.29 \quad \Rightarrow \quad \left|V_{tb}^{3\text{gen}}\right| = 0.99 \pm 0.15 \;\; (\text{stat} \oplus \text{syst})$$

$$\left|V_{tb}^{3\text{gen}}\right| > 0.80 \;\; (90\% \text{ CL}) \quad \left|V_{tb}^{3\text{gen}}\right| > 0.76 \;\; (95\% \text{ CL}).$$

If we relax the three generation unitarity constraint on $V_{tb}$ (and only on $V_{tb}$, not on the other elements in the CKM matrix), then the lower limit becomes:

$$\left|V_{tb}^{3\text{gen}'}\right| > 0.055 \;\; (90\% \text{ CL}) \quad \left|V_{tb}^{3\text{gen}'}\right| > 0.048 \;\; (95\% \text{ CL}).$$

To obtain the 3gen′ result, additional assumptions had to be made to replace the lost unitarity constraint on $V_{tb}$. The assumptions used by CDF are that the CKM matrix elements $V_{td}$ and $V_{ts}$ have the mean values from the SM three generation matrix: $V_{td} = 0.009$ and $V_{ts} = 0.040$.[2]

## 3   $V_{tb}$ and a Fourth Quark Generation

As shown above, the $t\bar{t}$ decay method for measuring $V_{tb}$ gives no information on the existence of a fourth quark generation unless we know $V_{td}$ and $V_{ts}$. But it is a possible fourth generation that we are interested in. A family of quarks beyond the three known ones has not been ruled out by experimental measurements, and is indeed quite possible or even favored in some extensions of the SM such as grand unified and string theories. If $V_{tb}$ is measured and found to be $\neq 1$, then there must be a fourth generation mixing with the third, or some other more exotic physics beyond the SM.

What do we know about a fourth generation so far? The LEP experiment ALEPH has searched for a down-type $b'$ quark produced in pairs from $Z$ bosons, and then decaying into any mode, and has set a limit of $m_{b'} > 46$ GeV at the 95% CL in 1990.[5] The CDF collaboration has searched for pair production of $b'$ quarks with charged current (CC) decay into a charm quark (skipping the heavy $t$ generation) and a lepton and neutrino. In 1992, they ruled out a $b'$ below 85 GeV with this decay mode.[6] It should be noted that CC decays are expected to be highly suppressed for a $b'$ with mass less than the top and less than a $t'$ quark. Finally, the DØ collaboration has recently published the result of a new search for $b'$'s with flavor changing neutral current (FCNC) decays into $b\gamma$ or $bg$. These modes are strongly favored for moderate values of $m_{b'}$. The 1997 result rules out a $b'$ with mass less than 96 GeV that has FCNC decays.[7]

Since the $t\bar{t}$ decay method will not give us model-independent information about a fourth quark generation, we need a new and more precise method for the next Tevatron run. Fortunately such a method exists.[8]

## 4   Measuring $V_{tb}$ Using Single Top Production

Top quarks can be produced singly via the electroweak interaction. There are two significant channels at the Tevatron: s-channel "$W^*$ production" with a cross section $\sigma\left(p\bar{p} \to t\bar{b} + \bar{t}b + X\right) = 1.0$ pb; and t- and u-channel "W-gluon fusion", with a cross section $\sigma\left(p\bar{p} \to tq + \bar{t}\bar{q} + X\right) = 3.0$ pb. These two cross sections are for a top quark of mass 170 GeV/$c^2$, at the upgraded Tevatron energy of $\sqrt{s} = 2.0$ TeV, and are next-to-leading order (NLO) results.[9,10] For the same conditions (top mass, collision energy), the resummed NLO $t\bar{t}$ cross section[11] is ~8.0 pb.

Since *tb* coupling is present in the single top production vertex, as well as in top decay, the measurement is straightforward. The central value of $V_{tb}$ is obtained directly from the measured cross section, which is proportional to the vertex function squared, and so to $|V_{tb}|^2$. The error on $V_{tb}$ is just the error on the measured single top cross section added in quadrature with the error on the theoretical calculation of that cross section, all divided by two.

*4.1   Errors on Single Top Cross Sections*

First I present the predicted theoretical errors on s-channel $W^*$ production $q'\bar{q} \to t\bar{b}$. The discussion follows arguments by Smith and Willenbrock.[9] Since this process originates from quarks in the proton, and not gluons to first order, the calculation uses well-defined parton distributions in the region $x \approx 0.1$, $Q^2 = q^2_{W^*}$, where $q^2_{W^*}$ is the mass of the virtual *W* boson. Therefore the parton distribution functions add little to the overall error, and we take their contribution as 2%, based on the difference between CTEQ3M and MRS(A'). Additionally, this contribution will be pinned down directly by measuring the ratio of the cross sections of s-channel single top to the process $q'\bar{q} \to l\bar{\nu}$, which also proceeds via a virtual *W* boson. Next, because the QCD corrections have already been calculated at NLO,[9] the contribution from the factorization scale uncertainty is only 4%. There are no one-loop interferences between the initial and final states to complicate things, because of the colorless s-channel *W* boson between. These arguments and calculations lead to:

$$W^* \text{ production} \qquad \delta\sigma_{\text{theory}}\left(p\bar{p} \to t\bar{b} + X\right) \approx 4.5\%.$$

Now I move on to discuss t-channel single top production, $q'g \to tq\bar{b}$ and $q'b \to tq$, and the theoretical errors. This mode is more difficult to understand and calculate, because it starts mainly from gluons in the proton, and not from valence quarks. The uncertainty on the gluon distribution is not well quantified and I assume a 20% error from this source. The recent NLO calculation of Stelzer, Sullivan, and Willenbrock[10] estimates a contribution of 5% from the *b* distribution factorization scale, and I include this too. The resulting error on the cross section, with a large uncertainty from the gluon distribution error, is:

$$W\text{-gluon fusion} \qquad \delta\sigma_{\text{theory}}\left(p\bar{p} \to tq + X\right) \approx 21\%.$$

Finally I discuss predictions of experimental errors on the measured single top cross sections.[12] Precise measurements of single top production will be made at future runs of the Tevatron collider by the DØ and CDF experiments. The next run, known as Run 2, will be from 1999–2002, and the upgraded detectors will collect ~2 fb$^{-1}$ of integrated luminosity each. In order to avoid making a statistics-limited measurement of $V_{tb}$, both types of single top production will be used together. The events will be identified using one $b$ tagged jet (mainly the central highly energetic $b$ from the top quark decay). Run 3 will be from 2003–2007. It is proposed that both the DØ and CDF detectors are upgraded to be able to take data at very high luminosities, and that the Tevatron collider be gradually upgraded throughout this period to increase the deliverable luminosity. The goal of Run 3 is for each experiment to collect ~30 fb$^{-1}$ of data. The most precise method for measuring $V_{tb}$ in Run 3 is to identify just $W^*$ single top production,[13] as the theory error is smaller than for t-channel single top production, and it will be comparable to the experimental error from such large data sets. We will be able to separate s-channel single top events from other single top modes and from background by requiring two $b$ tagged jets.

Using results from the TeV-2000 report[14] for signal and background acceptances, and the errors on these and other quantities such as the luminosity, we obtain estimates of the event yields with $m_t = 170$ GeV/$c^2$ for Runs 2 and 3, and hence predict the experimental error on the cross sections.[12] In Run 2, considering all single top production modes, we expect to find ~800 events (~1/3 $e$+jets and $\mu$+jets, ~2/3 background), and in Run 3, we anticipate ~1,500 events (~1/3 $W^*$ single top $e$+jets and $\mu$+jets, ~2/3 background). These predictions lead to the following experimental errors:

Run 2     $\delta\sigma_{\text{expt}}(p\bar{p} \to t + \bar{t} + X)$   $\approx$ 9% (stat) $\oplus$ 16% (syst) $\approx$ 18%

Run 3     $\delta\sigma_{\text{expt}}(p\bar{p} \to t\bar{b} + \bar{t}b + X)$   $\approx$ 6% (stat) $\oplus$ 6% (syst) $\approx$ 8.7%

## 4.2   Sensitivity to $V_{tb}$ for DØ and CDF

Using the estimates for the errors on the single top cross sections from the theoretical calculations and from predictions of event yields at future runs, it is now possible to obtain the sensitivity for measuring $V_{tb}$ at the upgraded Tevatron. The results are:

Run 2     $\delta|V_{tb}^{4\text{gen}}| = (21\% \text{ (theory)} \oplus 18\% \text{ (expt)})/2 = 14\%$

Run 3     $\delta|V_{tb}^{4\text{gen}}| = (4.5\% \text{ (theory)} \oplus 8.7\% \text{ (expt)})/2 = 5\%$.

In the high statistics limit, the precision on $V_{tb}$ is limited by the error on the integrated luminosity, which is difficult to measure well. For these results I have used a value of 5% which has been achieved in Run 1. A new measurement technique will have to be developed for this to be improved.

I would like to emphasize that the measurement of $V_{tb}$ obtained from the single top cross section is not dependent on making any particular assumptions about whether the SM holds (that there are three quark generations or whether unitarity applies), or what values $V_{td}$ and $V_{ts}$ have. This method yields a meaningful measurement of $V_{tb}$ which will give us information about the existence of a fourth quark generation if $V_{tb}$ is found to deviate significantly from near unity.

## 5 Comparison with Future Measurements at Other Facilities

Other accelerator facilities will also be able to make a model-independent measurement of $V_{tb}$. How does the predicted sensitivity at the upgraded Tevatron compare? Is it competitive?

As shown by Stelzer and Willenbrock,[13] the $W^*$ single top production method will not work at the Large Hadron Collider (LHC) because the backgrounds from other single top modes and from $t\bar{t}$ will be too high. The s-channel single top cross section does not rise very fast as $\sqrt{s}$ increases; the process has a quark and antiquark in the initial state, and there are no valence antiquarks at a $pp$ machine. For $\sqrt{s} = 14$ TeV, $m_t = 170$ GeV$/c^2$, CTEQ3M and $Q^2 = m_t^2$, $\sigma_{LO}(pp \to t\bar{b} + \bar{t}b + X) = 10$ ($t\bar{b}$) + 9 ($\bar{t}b$) = only 19 pb.[12] It might be possible to use t-channel single top at the LHC, since the statistics will be much better: $\sigma_{LO}(pp \to tq + \bar{t}\bar{q} + X) = 126$ ($tq$) + 79 ($\bar{t}\bar{q}$) = 205 pb.[12] However, the gluon distribution function will have to be measured with rather good precision to make a better than 5% measurement on $V_{tb}$. Detailed studies of this possibility have not been carried out.

At a future $e^+e^-$ collider (NLC), several methods have been considered for measuring $V_{tb}$. A $\sqrt{s} \approx 340$ GeV machine scanning the $t\bar{t}$ threshold can measure $V_{tb}^{4\text{gen}}$ to 4% precision with 100 fb$^{-1}$ of data.[15] If an NLC has $\sqrt{s} = 1$ TeV, then measuring the single top cross section using several ten's of fb$^{-1}$ will also lead to a 4% error.[16] Finally, using the single top process $e^-\gamma \to \bar{t}b\nu$ at $\sqrt{s} = 1$ TeV with 40 fb$^{-1}$ will provide a 5% precision on $V_{tb}^{4\text{gen}}$.[17]

Therefore, future measurements at other experimental facilities will not be any better for measuring $V_{tb}$ than the DØ and CDF experiments at the upgraded Tevatron at Fermilab, where it can be measured to a precision of 5%.